\documentclass[11pt]{article}
\usepackage{epsfig,epsf,latexsym,cite,times}

\title{ {\Large\sc Elastic moduli of model random three-dimensional
closed-cell cellular solids\footnote{Submitted to {\em Acta Materialia}}}}
\author{ {\normalsize A. P. ROBERTS$^{1,2}$ and E. J. GARBOCZI$^{1}$}
\\ {\small $^1$Building Materials Division,} \\
{\normalsize National Institute of Standards and Technology,
Gaithersburg, MD 20899, USA } 
\\ {\small $^2$Centre for Microscopy and Microanalysis,} \\
{\normalsize University of Queensland, St.\ Lucia, Queensland 4072,
Australia} \\ \\
}
\date{ {\normalsize April 4, 2000}}

\begin{document}
\maketitle

\begin{abstract}
Most cellular solids are random materials, while practically all
theoretical results are for periodic models. 
To be able to generate theoretical
results for random models, the finite element method (FEM) was used to study 
the elastic properties of solids with a closed-cell cellular structure.
We have computed the density ($\rho$) and microstructure dependence of the
Young's modulus ($E$) and Poisson's ratio (PR) for several different 
isotropic random models based
on Voronoi tessellations and level-cut Gaussian random fields. The effect
of partially open cells is also considered.
The results, which are best described by a power law
$E\propto\rho^n$ ($1 < n <2$), show the influence of randomness
and isotropy on the properties of closed-cell cellular
materials, and are found to be in
good agreement with experimental data.

\vspace{3mm}

\end{abstract}

\section{Introduction}

Manufactured cellular materials have been developed for a range of
applications~\cite{Gibson88}
(e.g., insulation, light-weight reinforcement),
and their natural counterparts (e.g. wood) have a 
cellular structure that optimizes performance for their particular 
requirements.
The useful properties of cellular solids depend on the
material from which they are made,
their relative density, and their internal geometrical structure. 
It is important to link the physical properties of
cellular solids to their density and complex microstructure, in order to
understand how such properties can be optimized for a given
application. Many studies have focussed on the elastic response of
periodic materials.  Equally important is the effect of
disorder (e.g., isotropy), and the interaction between cells on
a mesoscopic scale, as most real cellular solids are not periodic.
In this paper we study model isotropic cellular solids
at scales ($\approx$ 100 cells) where these effects can be probed.

At low densities, experimental results indicate that the Young's modulus
($E$) of cellular solids is related to their density ($\rho$) through the
relation~\cite{Gibson88}:
\begin{equation}
\frac E{E_s}= C \left( \frac{\rho}{\rho_s} \right)^n = C p^n
\label{basicE}
\end{equation}
where $E_s$ and $\rho_s$ are the Young's modulus and density of the solid
skeleton and $p=\rho/\rho_s$ is the reduced density. The constants
$C$ and $n$ depend on the microstructure
of the solid material.  The value of $n$ generally lies
in the range $1 < n < 4$, giving a wide range of properties at
a given density. For closed-cell foams, experimental studies indicate
that $1 < n < 2$.
The complex dependence of $C$ and $n$ on microstructure is
not well understood, and this remains a crucial problem in the ability
to predict and optimize the elastic properties of cellular solids.
At the local or cellular scale, important variables include the
cell character (e.g.~open or closed), the geometrical arrangement of
the cell elements (e.g.~angle of intersection), and the shape of
the cell walls (e.g.~curvature).
At a larger scale, the geometrical arrangement of the
cells is also crucial. The values of both $C$ and $n$ will depend on
whether the material is periodic or disordered.

Analysis of simple models shows that three basic mechanisms of deformation
are important for closed-cellular solids. If the cell walls are much
thinner than the cell edges, the deformation is governed by edge-bending.
In this case, $E$ varies quadratically with density ($n=2$),
and can be described by results for open cellular solids~\cite{Gibson88}.
If cell-wall bending is the mechanism of
deformation, Gibson and Ashby~\cite{Gibson82} have shown that $E$
should vary cubicly ($n=3$) with density. However, the fact that
$1 < n < 2$ indicates that cell-wall stretching ($n=1$) is actually
the dominant behaviour~\cite{Green85,Gibson88}.

\begin{figure}
\centering \epsfxsize=6.9cm\epsfbox{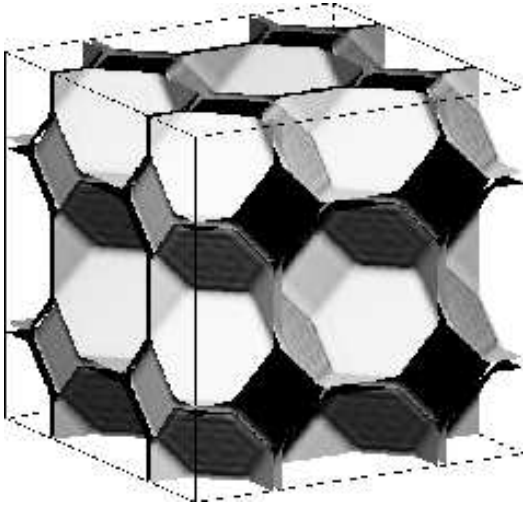} 
\caption{
The tetrakaidecahedral model. The model corresponds to the
Voronoi tessellation of a body-centered cubic lattice.
\label{tetra3D}}
\end{figure}

The ``tetrakaidecahedral'' foam model, in particular, has been the
subject of many recent
studies~\cite{Renz82,Simone98,Gren_shape99,Gren_perfect99,Mills_foam99}.
The cells of the model uniformly partition space, and are defined by
truncating the corners of a cube giving eight hexagonal and
six square faces (Fig.~\ref{tetra3D}).
The foam has a relatively low anisotropy~\cite{Simone98}
($E$ varies by less than 10 \% with direction of loading),
and is thought to be a good model of isotropic cellular solids.
In all cases, $E$ was found to decrease
linearly with density ($n=1$). However, real materials 
exhibit a larger dependence of $E$ on density ($n>1$),
indicating that periodic models do not capture salient features
of foam microstructure. It is likely that the random disorder
is responsible, and it is important to study its influence
on the properties of cellular solids.

There have been several recent studies of the effect of disorder in 
cellular solids. For two-dimensional models, variation in cell-shape
leads to a variation of 4 \% to 9 \% in elastic
properties~\cite{Silva97_elastic}, while
deletion of 5 \% of cell struts decreased the modulus by
35 \%~\cite{Silva97_compress}. 
Similar effects of `imperfections' were seen in
spring lattices, which have some similarities to foams~\cite{EJGSpring1}. 
In three-dimensions, Grenestedt~\cite{Gren_shape99} showed that disorder
decreased the Young's modulus of the
tetrakaidecahedral
foam (with 16 cells) by 10 \%. However, only the pre-factor of
Eq.~(\ref{basicE}) was affected; the scaling exponent remained constant ($n=1$).
Grenestedt~\cite{Gren_wavy98} has
also estimated the effect of ``wavy-imperfections'' on the stiffness
of a cube with closed cell walls. If the wave-amplitude was five times the
cell-wall thickness, the stiffness decreased by 40 \% compared to
the case of flat faces. 

From the foregoing discussion it is clear that more
complex, three-dimensional random models are necessary
to improve predictions for cellular solids~\cite{Roberts95b}. 
There are two main problems in studying random models. First, 
a sufficiently accurate model
of the microstructure must be developed. And second, the properties
of the model must be accurately evaluated. We emphasize that there are no exact
analytical calculations available for general random materials, 
so that numerical
methods become necessary.

In this paper we use a finite element method (FEM)~\cite{Garboczi95a}
to estimate the elastic properties of model cellular solids
over a range of densities.
The models are generated using tessellation
methods~\cite{StoyanSG} and level-cut random field
models~\cite{Berk87}. 
The Young's moduli of the models can be described in
terms of simple two parameter relations [e.g.,~Eq.~(\ref{basicE})
in the low density limit].  The results demonstrate the effect of
microstructure, isotropic disorder, and finite density
on the elastic properties of cellular solids, including both Young's modulus and
Poisson's ratio.
Apart from the small numerical errors in the finite
element method, $10$ \% or less, the results are exact for each 
of the models. Therefore,
the results can be used to predict the properties of cellular solids
if their structure is similar to one of the models, or be used
to accurately interpret experimental data.

\section{Prior results}
\label{prior}

In this section we discuss prior results for closed-cell foams. The
results illustrate, and attempt to quantify, the basic mechanisms
of deformation. We compare the results to our FEM results in subsequent
sections to illustrate the effect of disorder in multi-cellular
models.

Christensen~\cite{ChristensenLow86} has derived a result for
a closed cell material comprised
of randomly located and isotropically oriented large intersecting
thin plates. The results are,
\begin{equation}
\frac EE_s=
\frac{2 (7-5\nu_s)}{3(1-\nu_s)(9+5\nu_s)} \frac \rho\rho_s;\;\;
\nu= \frac{1+5\nu_s}{9+5\nu_s}.
\label{thyCC}
\end{equation}
where the subscript ``s'' indicates the solid phase. 
The linear dependence ($n=1$) of modulus on density
is typical for cellular materials with `straight-through' elements. In
this case, cell-wall stretching is the only mechanism of deformation.

Analysis of more complex closed cell foams is very difficult,
but computational results
~\cite{Gren_shape99,Gren_perfect99,Mills_foam99,Renz82,Simone98}
have been obtained for the closed cell 
tetrakaidecahedral foam shown in Fig.~\ref{tetra3D}.
Simone and Gibson~\cite{Simone98} recently found that the Young's modulus
is nearly equal (within 10 \%) for loading in the
$\langle100\rangle$, $\langle111\rangle$ and $\langle110\rangle$ directions.
For the density range $0.05 < \rho/\rho_s <0.20$, their results for
the $\langle100\rangle$ direction were fitted with the formula
\begin{equation}
\label{thySimone}
\frac {E_{100}}{E_s} \approx 0.315 \left( \frac\rho\rho_s \right) +
                      0.209 \left( \frac\rho\rho_s \right)^2
\end{equation}
which compares well with the earlier result
${E_{100}}/{E_s}\approx 0.33 (\rho/\rho_s)$~\cite{Renz82}.
For the case where the face thickness is 5 \% of the edge thickness,
Mills and Zhu~\cite{Mills_foam99} found
${E_{100}}/{E_s}\approx 0.06 (\rho/\rho_s)^{1.06}$ 
in the density range $0.015<\rho/\rho_s < 0.1$.

Gibson and Ashby have proposed the semi-empirical formulae
\begin{equation}
\label{thyGA_C}
\frac EE_s \approx \phi^2
\left( \frac\rho\rho_s  \right)^2 + (1-\phi) \frac\rho\rho_s;
\;\; \nu \approx \frac 13
\end{equation}
where $\phi$ is the fraction of solid mass contained in the
cell-edges (the remaining fraction $1-\phi$ is in the cell faces).
Gas trapped in the cells can also increase the stiffness, but this
effect is usually negligible~\cite{Gibson88}.
The first term of Eq.~(\ref{thyGA_C})
accounts for deformation in the cell edges. Note that 
the case $\phi$=1 corresponds to a commonly used
semi-empirical formula for open-cell solids, i.e.\
Eq.~(\ref{basicE}) with $C=1$ and $n=2$.
The second term corresponds to stretching deformation in the
cell faces.
The result provides good agreement with data for closed-cell foams
when $0.6 \leq \phi \leq 0.8$~\cite{Gibson88}. In part, the implied relatively
high cell-edge fractions can be attributed to the fact that
surface tension forces drive mass out of the cell walls
into the edges. However, in some foams, the cell faces are relatively
thick, giving $\phi$=0.01 -- 0.07~\cite{Mills_foam99} (note that these
authors report the fraction of total mass in the
{\em faces} which corresponds to $1-\phi$) and we expect
Eq.~(\ref{thyGA_C}) to overestimate the measured values.

There are also several kinds of exact bounds that have been derived for
the elastic properties of composite materials~\cite{TorqRev91}. If the
properties of each phase in a composite are not too dissimilar, the
bounds can be quite restrictive.  For porous materials, however,
the bounds on Poisson's ratio are no more restrictive than the
range guaranteed by the non-negativity of $K$ and $G$ for isotropic materials
($-1\leq \nu \leq 0.5$), and
the lower bound on $E$ reduces to zero.
Nevertheless, the upper bound on $E$
is sometimes found to provide a reasonable approximation of the actual
property.

The most commonly applied bounds for isotropic composites are
due to Hashin-Shtrikman~\cite{HSelas}.
These bounds can be evaluated if the elastic
properties and volume fraction of each phase are available.
The upper bound
$E_u$ is
\begin{eqnarray}
\frac{E_u}{E_s} &=& \frac{p}{1+C_H(1-p)};\;\; \nu_u=0.5 \\
C_H&=&\frac{(1+\nu_s)(13-15\nu_s)}{2(7-5\nu_s)}.
\label{HSthy}
\end{eqnarray}
Note that $C_H(\nu_s=0.2)=1$, and $11/12 \leq C_H < 1.006$
for $\nu_s>0$ (the maximum occurring near $\nu_s=0.27$).
Therefore as $(\rho/\rho_s)\to0$
$E/E_s \approx \frac12 (\rho/\rho_s)$.
In order to improve the bound, it is necessary
to know the $N$-point correlation functions of the
composite~\cite{TorqRev91,Milton82b}. 
These functions are generally only available for certain models to
order $N=3$. In this case, the bounds are referred to as
3-point bounds.

It is interesting to compare the bound with the formulae reported above.
It is simple to show that Eqs.~(\ref{thyCC}) and ~(\ref{HSthy})
are identical as $\rho/\rho_s\to0$. This indicates that, to the
accuracy of Christensen's approximation, randomly oriented
straight-through plates provide an optimally stiff microstructure.
Also note that the semi-empirical formula given in Eq.~(\ref{thyGA_C})
actually violates the bound if more that half the solid material resides
in the cell faces ($\phi < 0.5$).

\section{Finite Element Method}
\label{fem}

The finite element method uses a variational formulation of the
linear elastic equations, and finds the solution by minimizing
the elastic energy via a fast conjugate gradient method.
The FEM we use has been especially adapted
for periodic rectangular parallelepiped digital images 
(although they can be used on non-periodic
images).  The algorithm handles only linear elasticity at
present, although it is not in principle restricted to only
linear elasticity.

Each pixel, in 3-D, is taken
to be a tri-linear finite element~\cite{Cook}.
For random materials, it is much easier to mesh using
the pixels of the digital image lattice
rather than a collection of beams, plates, etc.
The digital image is assumed to have periodic boundary conditions.
A strain is applied, with the average stress or the average elastic
energy giving the effective elastic moduli~\cite{TorqRev91,Hashin83}.
Details of the theory and copies of the actual programs used are
reported in the papers of Garboczi \& Day~\cite{Garboczi95a}
and Garboczi~\cite{Eds_manual}. 

Given a digital microstructure, the finite element method
provides a numerical solution of the elasticity equations.
The accuracy is only limited by the finite number of pixels which
can be used (around $10^6$ in this study). Preliminary
studies indicated that about 100 cells are necessary to properly
simulate the macroscopic properties of a cellular solid (which
may have many thousands of cells). We generally calculated
the properties of five samples at each density
and report an average value. The
statistical uncertainty in the results is estimated to be
less than 10 \%.
Note that if a foam is regular and periodic, 
just one measurement on a unit cell is sufficient. 

A potentially greater source of error occurs in the finite element
method when there are insufficient pixels in a solid region to
correctly model continuum elasticity. 
A useful method of estimating these discretization errors
is to compute the properties of regular periodic foams,
since in these models there are no sources of statistical error, and
there are exact solutions to which to compare the numerical results.
The foams we consider have cubic symmetry, which
means that the direction dependent elastic
properties can be characterized by three independent
constants, $C_{11}$, $C_{12}$ and $C_{44}$, of the Hooke's law stress-strain
tensor~\cite{ChristensenBook}.
For loading along the $x$-axis
(ie.\ one of the axes of symmetry) the Young's modulus
and Poisson's ratio are
\begin{eqnarray}
E_{100}&=&\frac{(C_{11}-C_{12})(C_{11}+2C_{12})}{(C_{11}+C_{12})} \\
\nu_{12}&=&\frac{C_{12}}{(C_{11}+C_{12})}.
\end{eqnarray}
The bulk modulus is actually independent of direction and given
by $K=E_{100}/3(1-2\nu_{12})$, and the anisotropic shear modulus (for shearing
parallel to a symmetry plane) is just $C_{44}$.
The finite element codes evaluate the $C_{ij}$
directly, but for simplicity we report the engineering constants.

To check the effect of resolution for the finite element method
we measured the Young's modulus ($E_{100}$) of two tetrakaidecahedral models
with edges of thickness 4 pixels and 8 pixels, respectively.
We found virtually no difference ($<$ 1 \%) in $E_{100}$ 
indicating that the discretization errors are quite small. However,
the absolute value is 15 \% greater than that estimated by Simone
and Gibson~\cite{Simone98} using a specially tailored finite-element
grid. Since we have found our FEM to be accurate for many other
test cases, the origin of the discrepancy is unclear. In
related studies, we have found discretization errors of around 10 \%,
and we assume this will be true for the random models studied here.

\section{Elastic properties of model cellular solids}
\subsection{Voronoi tessellations}
\label{elastic_vt}

The most common models of cellular solids are generated
by Voronoi tessellation of distributions of `seed-points'
in space.  Around each
seed there is a region of space that is closer to that seed
than any other. This region defines the cell of a Voronoi
(or Dirichlet) tessellation~\cite{StoyanSG}. 
The Voronoi tessellation can also be obtained~\cite{StoyanSG} by allowing
spherical bubbles to grow with uniform velocity from each of the seed points.
Where two bubbles touch, growth is halted at the contact point, but allowed
to continue elsewhere. In this respect the tessellation is similar to
the actual process of liquid foam formation~\cite{Schul_foam97}.
Of course physical
constraints, such as minimization of surface energy, will also play
an important role. Depending on the properties of the liquid and the
processing conditions, the resultant solid foam will be comprised
of open and/or closed cells.

It is worth noting that the tessellation of the BCC array
(the tetrakaidecahedral cell model discussed above) is a reasonable
approximation to the foam introduced by Lord
Kelvin~\cite{Weaire94,Gren_perfect99}.
The cells of the Kelvin foam are uniformly shaped, fill space, and
satisfy Plateau's law of foam equilibrium
(three faces meet at angles of 120$^o$, and four struts join at 109.5$^o$).
In order for this to be true, the faces and edges are slightly
curved~\cite{Weaire94}, unlike those of the tetrakaidecahedral cell model.

\begin{figure}[t!]
\begin{minipage}[t!]{1.\linewidth}
\begin{minipage}[bt!]{.5\linewidth} \noindent
\centering \epsfxsize=.85\linewidth\epsfbox{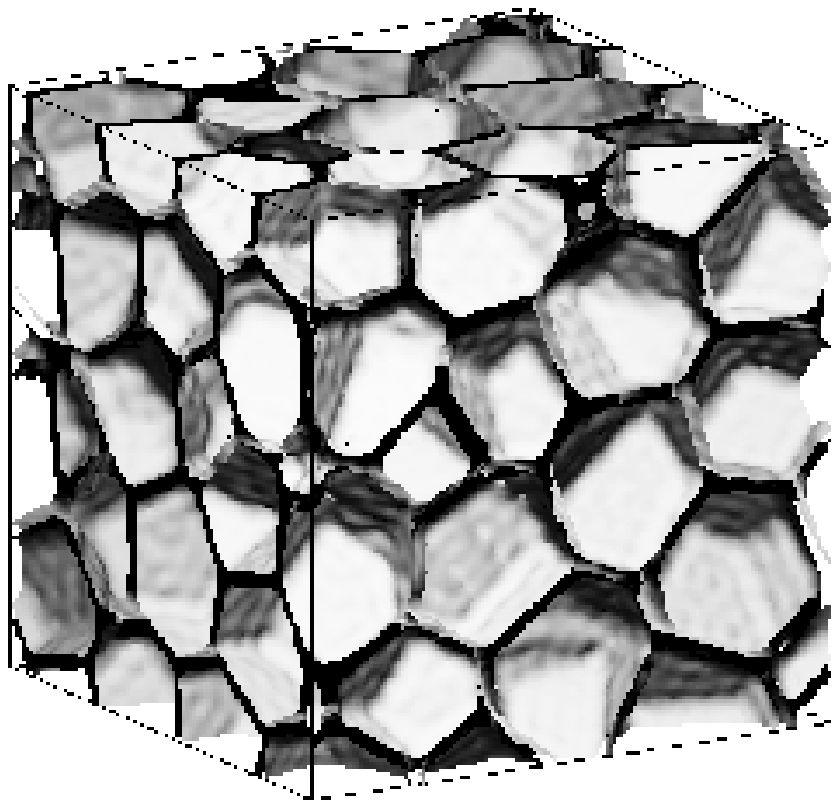}

Figure 2
\end{minipage}
\begin{minipage}[bt!]{.5\linewidth}
\centering \epsfxsize=.96\linewidth\epsfbox{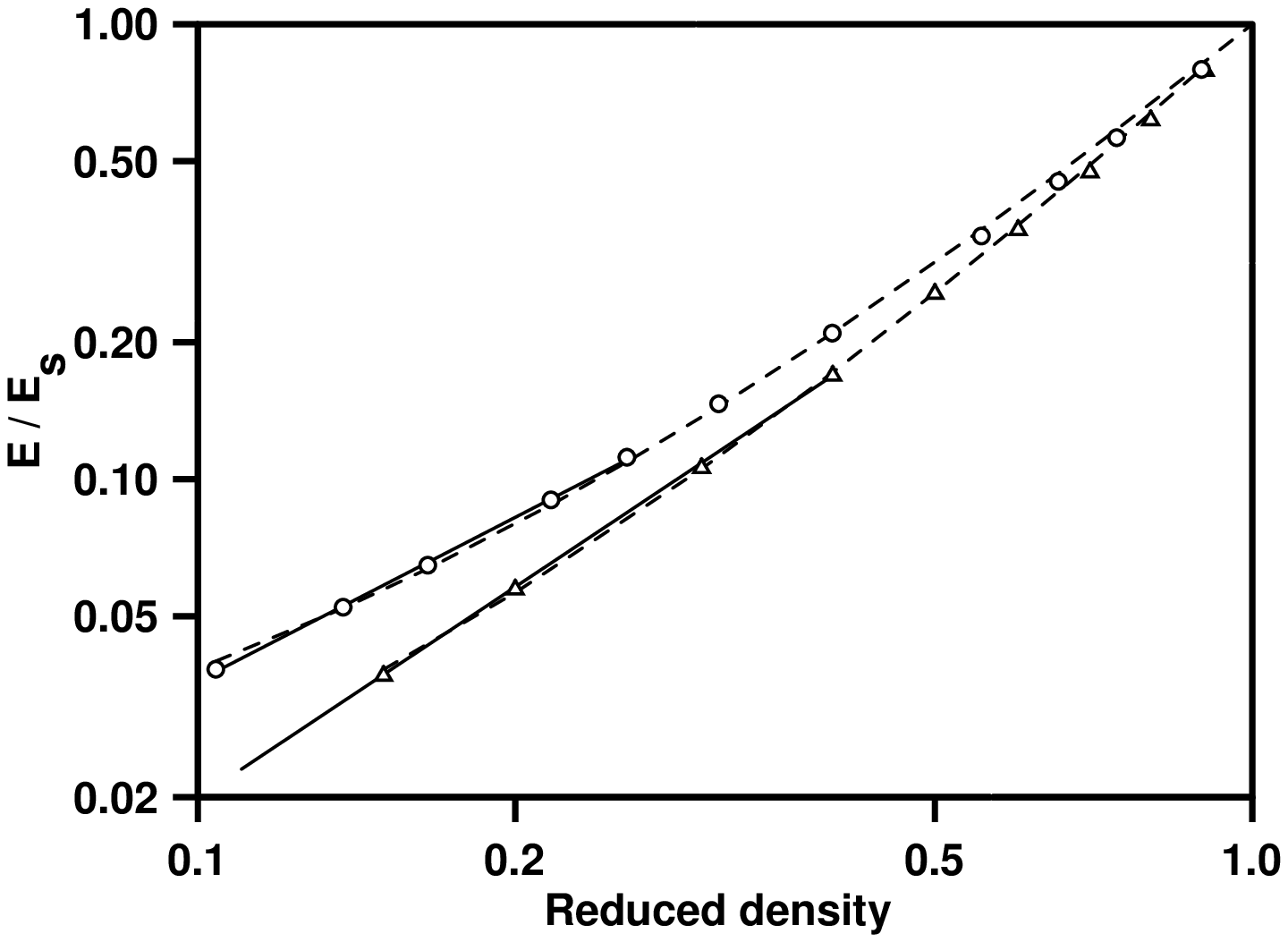}
Figure 3
\end{minipage}
\caption{The Voronoi tessellation model of a closed cell foam.
The reduced density is $\rho/\rho_s\approx 0.18$. The model shown has 63 cells, whereas the computations were performed
on samples with 122 cells.
\label{vt3D}}
\caption{The Young's ($\circ$) modulus of the Voronoi tessellation ($\circ$)
and Gaussian random field model ($\triangle$).
The solid lines are the empirical fits to the data given in
Eqs.~(\ref{vt_fit}) and (\ref{modU1_fit}), respectively. The dashed lines
are the fits to Eq.~(\ref{percfit}).
\label{vt_plot}}
\end{minipage}
\end{figure}

To generate foams with a roughly uniform
cell size we use 122 seed points corresponding to the center's of close-packed
(fraction 0.511) hard spheres in thermal equilibrium~\cite{TorqRev91}.
A pixel in the digital model is defined as belonging to a face 
if it is approximately equidistant from at least two sphere centers.
The density of the model is changed by varying the thickness
of the cell faces.
An illustration of the model (with only 63 cells)
is shown in Fig.~\ref{vt3D}.

\begin{table}
\caption{Properties of the closed cell Voronoi tessellation model.
Data at the two lowest densities were obtained for 3 realizations of
the 26 cell model, and the remainder were obtained for 5 realizations
of the 122 cell model. The relative density of the cell-edges
$(\rho/\rho_s)_o$ was obtained by deleting all the faces in the
closed-cell models. 
\label{tabvt}}
\begin{center}
\begin{tabular}{|c|c|c|c|c|}
\hline
$\rho/\rho_s$ & $E/E_s$ & $(\rho/\rho_s)_o$ & $\phi$ & $M$ 
\\
\hline
  0.104 & 0.038&0.016&0.15& 128 \\
  0.137 & 0.052&0.028&0.21& 96  \\
  0.165 & 0.065&0.045&0.27& 128  \\
  0.216 & 0.090&0.077&0.36& 96   \\
  0.255 &  0.11&0.l08&0.42& 80   \\
  0.312 &  0.15&0.16 &0.52& 64   \\
  0.400 &  0.21&0.24 &0.60& 48   \\
  0.553 &  0.34&0.35 &0.62& 64   \\
  0.655 &  0.45&0.46 &0.71& 64   \\
  0.744 &  0.56&0.56 &0.76& 64  \\
  0.895 &  0.80&0.80 &0.89& 64  \\
\hline
\end{tabular}
\end{center}
\end{table}

Using $M$=128 pixels in each direction to resolve the
structure, and a wall thickness of two pixels, the minimum density
obtainable (using 122 cells) was $\rho/\rho_s=0.16$. In order to examine
the stiffness at lower densities we also generated samples with 26 cells.
We found that foams of 26 and 122 cells had the same
stiffness (within 1 \%), indicating that finite size
effects are very small for the model. The results are given in
Table~\ref{tabvt} and
plotted in Fig.~\ref{vt_plot}.
In the low density limit the Young's modulus of the closed-cell model
can be described to within a 1.5 \% relative error by,
\begin{equation} \label{vt_fit}
\frac EE_s  = 0.563 \left(\frac{\rho}{\rho_s} \right)^{1.19} \rm{for}\;\;
0.1 < \frac\rho\rho_s < 0.3.
\end{equation}
This simple scaling relation cannot reproduce the high density behavior
($E\to E_s$ as $\rho \to \rho_s$) unless $C$ is fortuitously equal
to one. Rather than choosing a three- or four-parameter relation
to describe the full density range, we instead use the equation
\begin{equation} \label{percfit}
\frac E{E_s}= \left( \frac{ p-p_0 }{1 - p_0}\right)^m\;\;
(p=\frac \rho\rho_s), \end{equation}
which has been found to be useful for describing the properties at high
densities. 
With $m=2.09$ and $p_0=-0.140$ Eq.~(\ref{percfit}) describes the data
to within 4 \% for $0.15 < \rho/\rho_s < 1$.

\subsection{Effect of deleting faces}
\label{sec_del}

Depending on the physical conditions for foam formation, it is 
possible for the final foam to contain both open and closed cells.
It is relatively easy to delete cell-walls from the Voronoi
tessellation, which allows us to quantitatively investigate how the
presence of partially open cells degrades the foam stiffness.
In the Voronoi tessellation, a cell edge is defined by points
which are equidistant from three (or more) cell centers. If the
edges are retained but the cell walls are absent, an open-cell
Voronoi tessellation results. In a parallel study of open-cell
foams we have shown that the Young's modulus of the open-cell
foam is $E/E_s=0.93 (\rho/\rho_s)^{2.04}$ for $\rho/\rho_s < 0.5$,
in good agreement with scaling arguments.

\begin{figure}[t!]
\begin{minipage}[t!]{1.\linewidth}
\begin{minipage}[bt!]{.5\linewidth} \noindent
\centering \epsfxsize=.85\linewidth\epsfbox{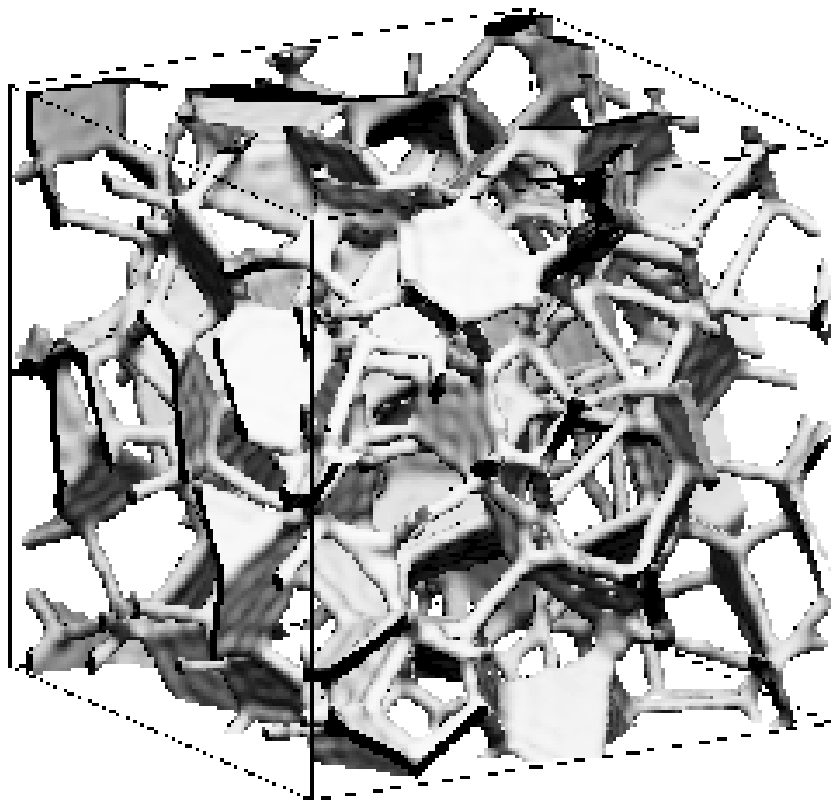}

Figure 4
\end{minipage}
\begin{minipage}[bt!]{.5\linewidth}
\centering \epsfxsize=.85\linewidth\epsfbox{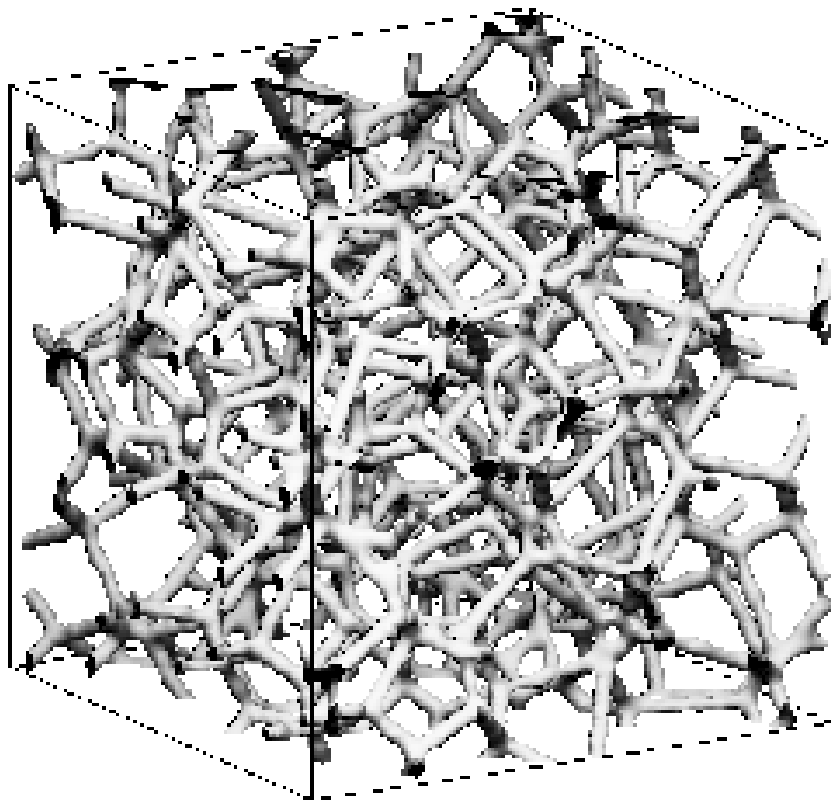}

Figure 5
\end{minipage}
\caption{The Voronoi tessellation model with 70 \% of the faces
deleted.
The reduced density is $\rho/\rho_s\approx 0.09$.
\label{vt3D70}}
\caption{The open-cell Voronoi tessellation model with reduced density
$\rho/\rho_s$=0.05. (All the faces have been deleted.)
\label{vte3D}}
\end{minipage}
\end{figure}

\begin{figure}[t!]
\begin{minipage}[t!]{1.\linewidth}
\begin{minipage}[bt!]{.5\linewidth} \noindent
\centering \epsfxsize=.95\linewidth\epsfbox{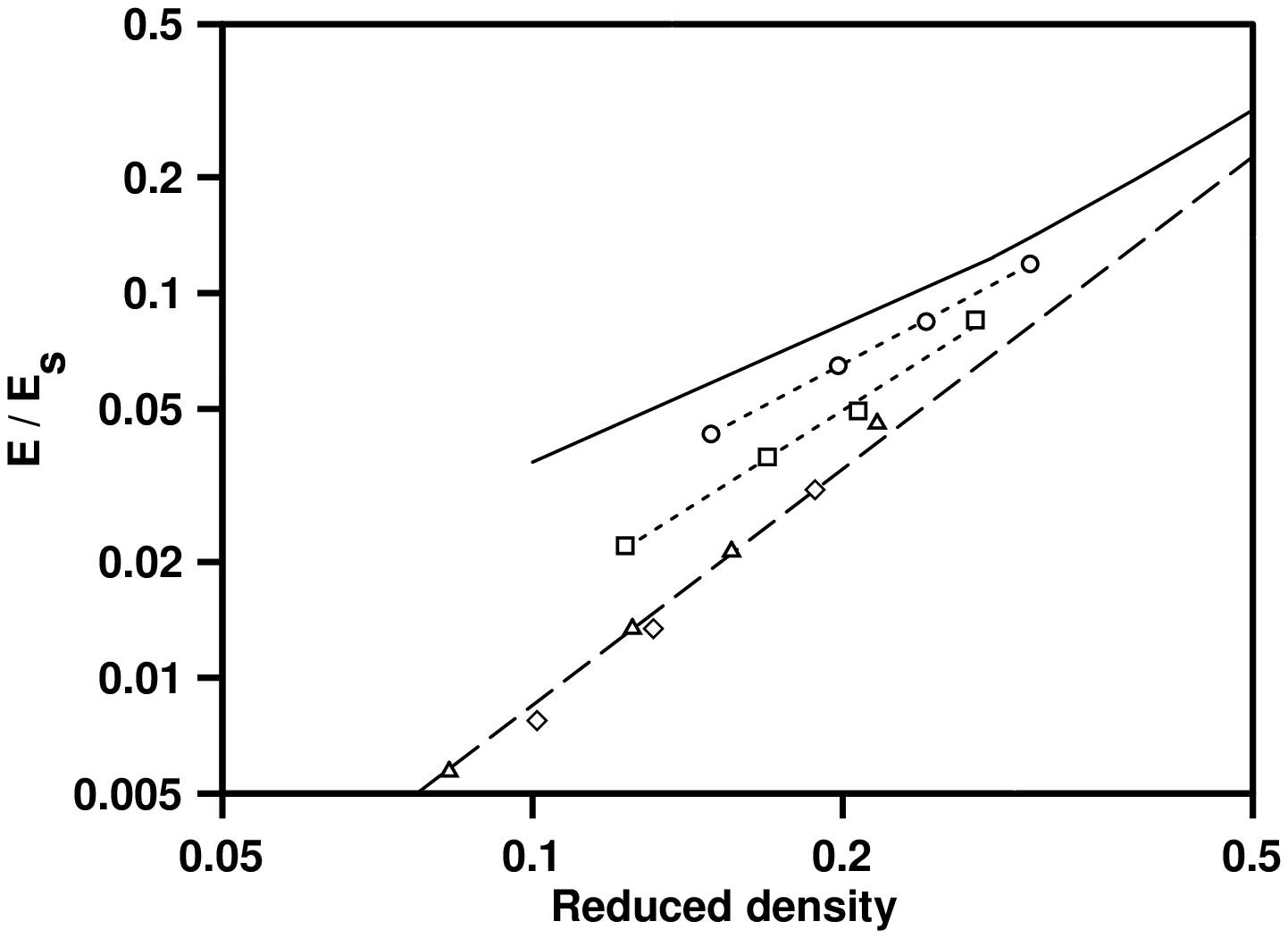}

Figure 6
\end{minipage}
\begin{minipage}[bt!]{.5\linewidth}
\centering \epsfxsize=.95\linewidth\epsfbox{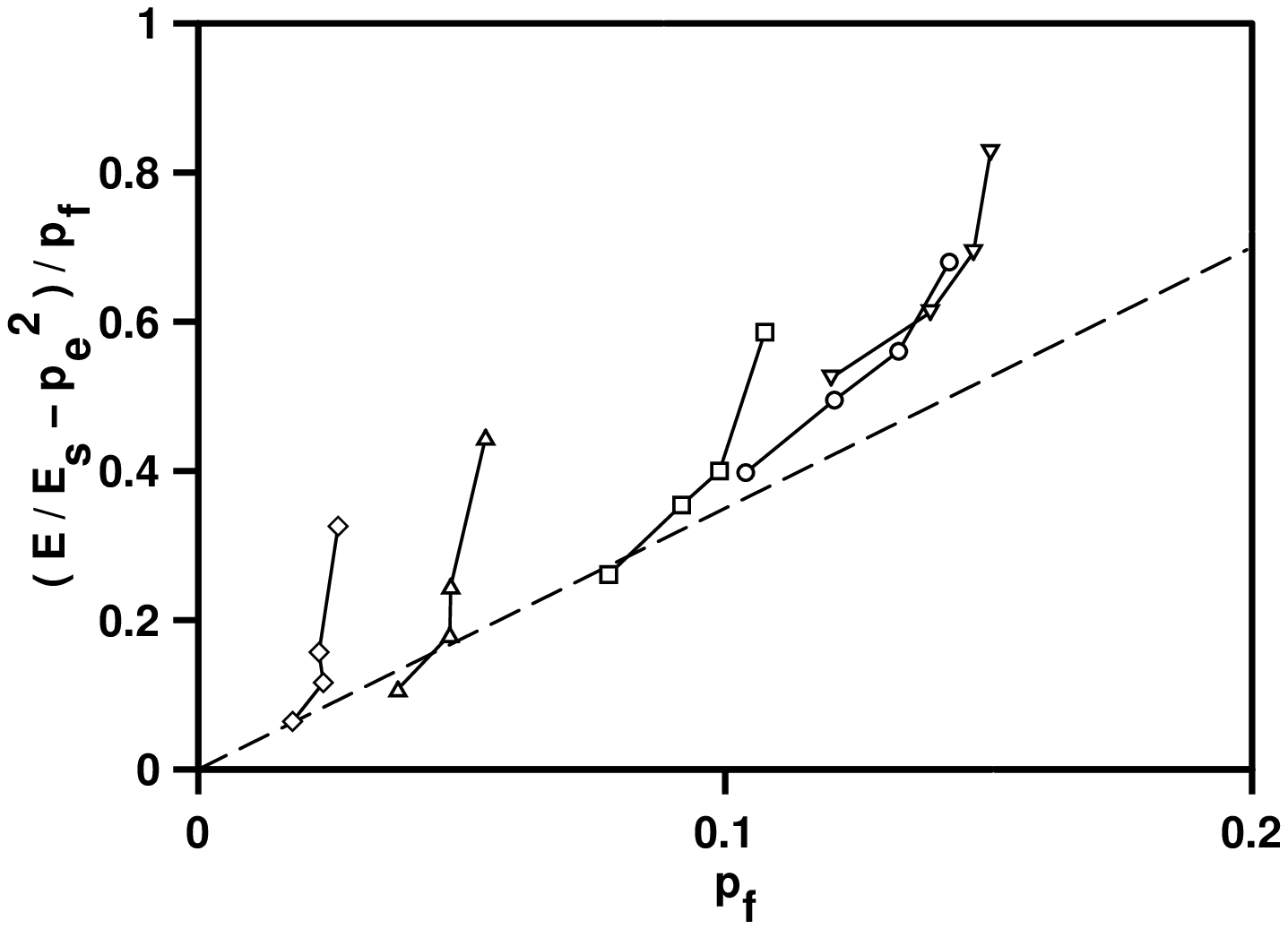}

Figure 7
\end{minipage}
\caption{The effect of deleting faces from the closed-cell Voronoi
tessellation. The symbols correspond to deletions of
20 \% ($\circ$), 40 \% ($\Box$),
70 \% ($\triangle$) and 85 \% ($\diamond$).
The extreme cases of no-deletions (------) and 100 \% deletions (--- ---)
are also shown. The dotted lines correspond to the empirical fits described
in the text.
\label{del_faces}}
\caption{If the modulus of the partially-open tessellation model
could be linearly resolved into a contribution from the cell faces and
edges (with fractions $p_{\rm f}$ and $p_{\rm e}$) then we would
expect $(E/E_s-p_{\rm e}^2)/p_{\rm f}$ to be independent of $p_{\rm f}$.  This
is seen not to be the case. The line has a slope of one.
The symbols correspond to deletions of 0 \% ($\bigtriangledown$),
20 \% ($\circ$), 40 \% ($\Box$),
70 \% ($\triangle$) and 85 \% ($\diamond$).
\label{extra_mass}}
\end{minipage}
\end{figure}

To test the intermediate
cases we consider tessellations with 20 \%, 40 \%, 70 \% and 85 \% of the
faces removed at random. The underlying edges of the open-cell
tessellation are left intact.  Examples of the microstructure are shown in
Figs.~\ref{vt3D70} and~\ref{vte3D} for the cases where 70 \% and 100 \%
of the cell walls have been deleted. The results are plotted in
Fig.~\ref{del_faces}. If 20 \% of the faces are deleted, we find
$E/E_s=0.64 (\rho/\rho_s)^{1.4}$ and when 40 \% of the faces are deleted,
the result is $E/E_s=0.76 (\rho/\rho_s)^{1.7}$. For 70 \% and above, the
Young's modulus follows the open-cell result.

It would be theoretically useful to partition the results for partial
deletion into a contribution from edge-bending and plate stretching, similar
to Eq.~(\ref{thyGA_C}). For each model, we can directly measure the
respective solid fractions $p_{\rm e}$ and $p_{\rm f}$ of the edges and faces,
with $p_{\rm e}+p_{\rm f}=\rho/\rho_s$. In the absence of
cell walls, the edges have a modulus of $E/E_s \approx p_{\rm e}^{2}$,
and the cell walls should contribute a term which depends linearly on
$p_{\rm f}$.  Thus, if the contributions can be linearly combined, a plot
of
$F(p_{\rm f})=(E/E_s - p_{\rm e}^{2})/p_{\rm f}$ vs.\ $p_{\rm f}$
should yield
a constant value. This is seen not to be the case in Fig.~\ref{extra_mass}.
Indeed $F(p_{\rm f})$ is
seen to increase nearly linearly with $p_{\rm f}$ indicating that the
additional contribution of the mass in the cell walls to the
Young's modulus approximately follows a quadratic law.
We conclude that it is not possible to describe the Young's modulus of
the partially open cell model in terms of a contribution of `edge-bending'
and `plate-stretching'. Clearly both mechanisms are active in deformation,
but they combine non-linearly. Our evidence suggests that the data can
be best represented by a power law with a non-integer exponent
$1 < n < 2$.

\subsection{Gaussian random fields}
\begin{figure}[t!]
\begin{minipage}[t!]{1.\linewidth}
\begin{minipage}[bt!]{.5\linewidth} \noindent
\centering \epsfxsize=.80\linewidth\epsfbox{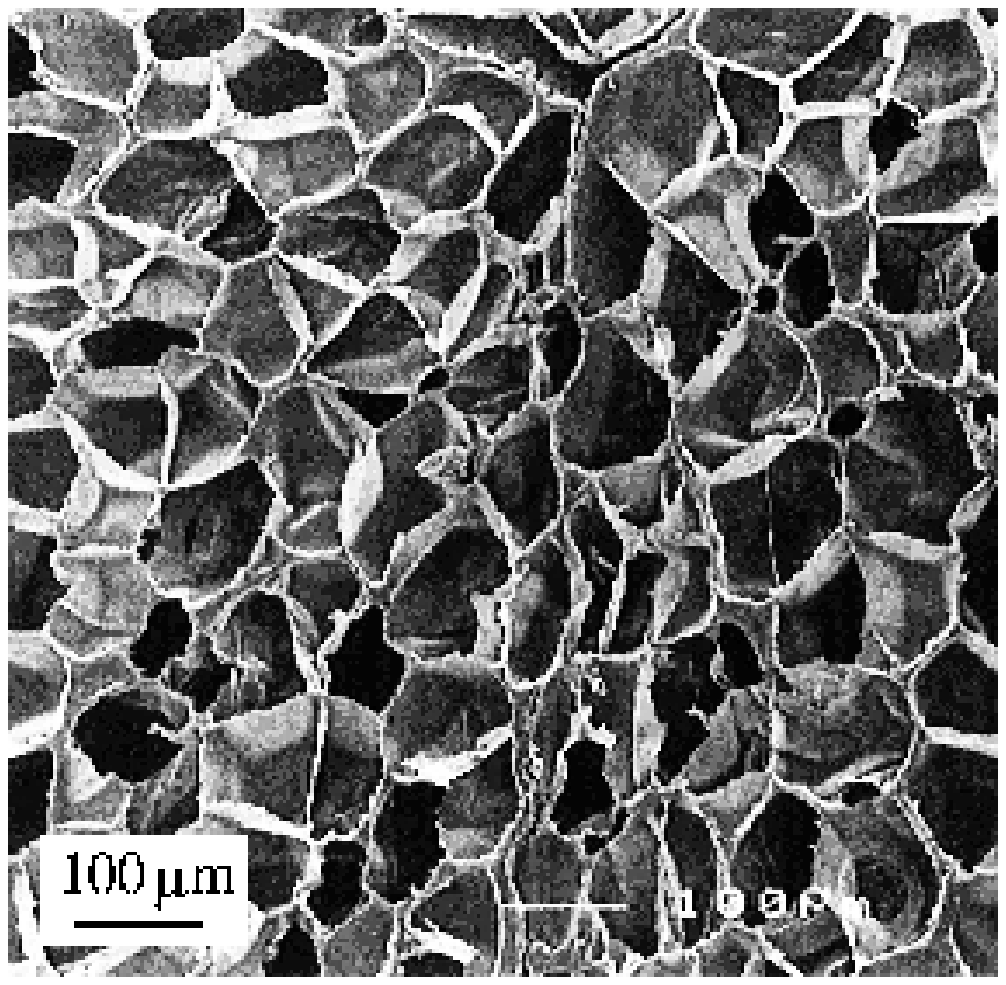}

Figure 8
\end{minipage}
\begin{minipage}[bt!]{.5\linewidth}
\centering \epsfxsize=.85\linewidth\epsfbox{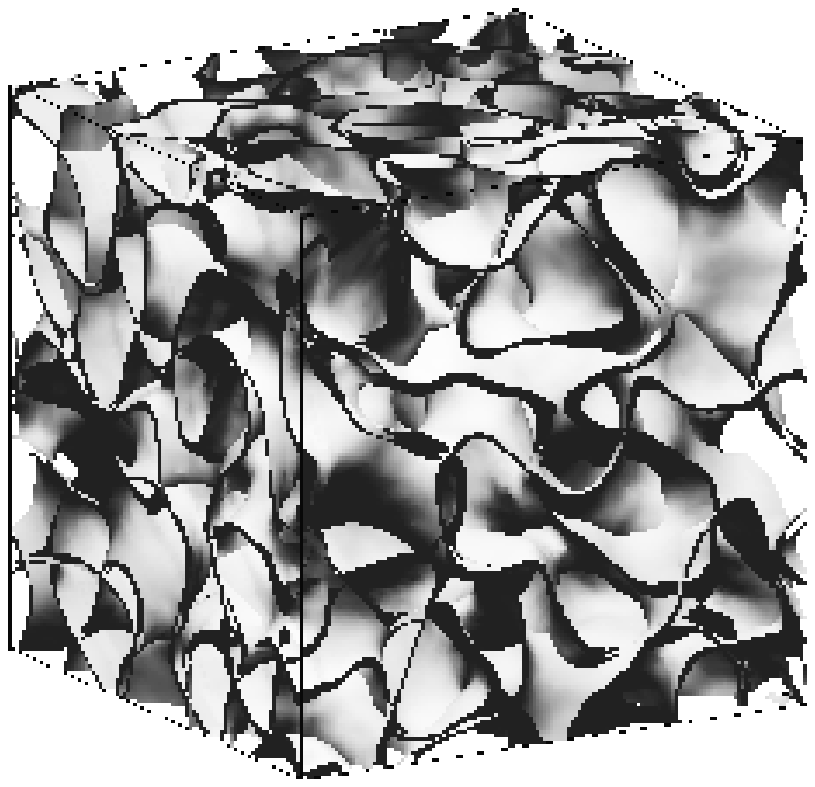}

Figure 9
\end{minipage}
\caption{Irregular cell shapes and sizes in polystyrene
foam~\cite{Roberts95b}.
\label{cc_sem}}
\caption{The closed-cell Gaussian random field model with
reduced density $\rho/\rho_s\approx 0.2$.
\label{modU1_3D}}
\end{minipage}
\end{figure}

The Voronoi tessellation has regular cells with perfectly flat walls.
However some cellular solids, such as the polystyrene sample shown in
Fig.~\ref{cc_sem},
have irregularly shaped cells which are likely to reduce their
stiffness~\cite{Gren_wavy98}.
Since it is difficult to include this type of randomness in
the Voronoi tessellation we consider a statistical model based on
Gaussian random fields (GRFs), which shows a large variation in cell
shapes and sizes.
To generate the model, one starts with a GRF
$y({\bf r})$ which
assigns a (spatially correlated) random number to each point in space.
A two-phase solid-pore
model~\cite{Berk87,Roberts96a} can be defined by letting the
region in space where
$-\beta < y({\bf r}) <\beta$ be solid, while the remainder
[$|y({\bf r})| \geq \beta$]  corresponds to the pore-space.
A closed-cell model can be
obtained from the model by forming the union set of two
statistically independent level cut GRF models~\cite{RobertsAero}.
Details for generating the models have been previously
described~\cite{Roberts_TAg99}.
An example is shown in Fig.~\ref{modU1_3D}. While not exactly
reproducing the structure of polystyrene, the model
is able to qualitatively mimic the closed cells and curved
walls seen in Fig.~\ref{cc_sem}.

The Young's modulus of the model can be described
to within a 2 \% relative error by,
\begin{equation} \label{modU1_fit}
\frac EE_s  = 0.694 \left(\frac{\rho}{\rho_s} \right)^{1.54} \rm{for}\;\;
0.15 < \frac\rho\rho_s < 0.4 
\end{equation}
in the low density regime, and Eq.~(\ref{percfit}) with $m$=2.30
and $p_0=-0.121$ for $0.15  < \rho/\rho_s < 1$ (relative error 3 \%).
The data and fitting curves are shown in Fig.~\ref{vt_plot}.
As in the case of the closed-cell Voronoi tessellation, the difficulty of
resolving the very thin cell walls prohibits lower densities from being
studied at present.

\section{Comparison of FEM results with prior results}
\label{compareT}

\begin{figure}[t!]
\begin{minipage}[t!]{1.\linewidth}
\begin{minipage}[bt!]{.5\linewidth} \noindent
\centering \epsfxsize=.95\linewidth\epsfbox{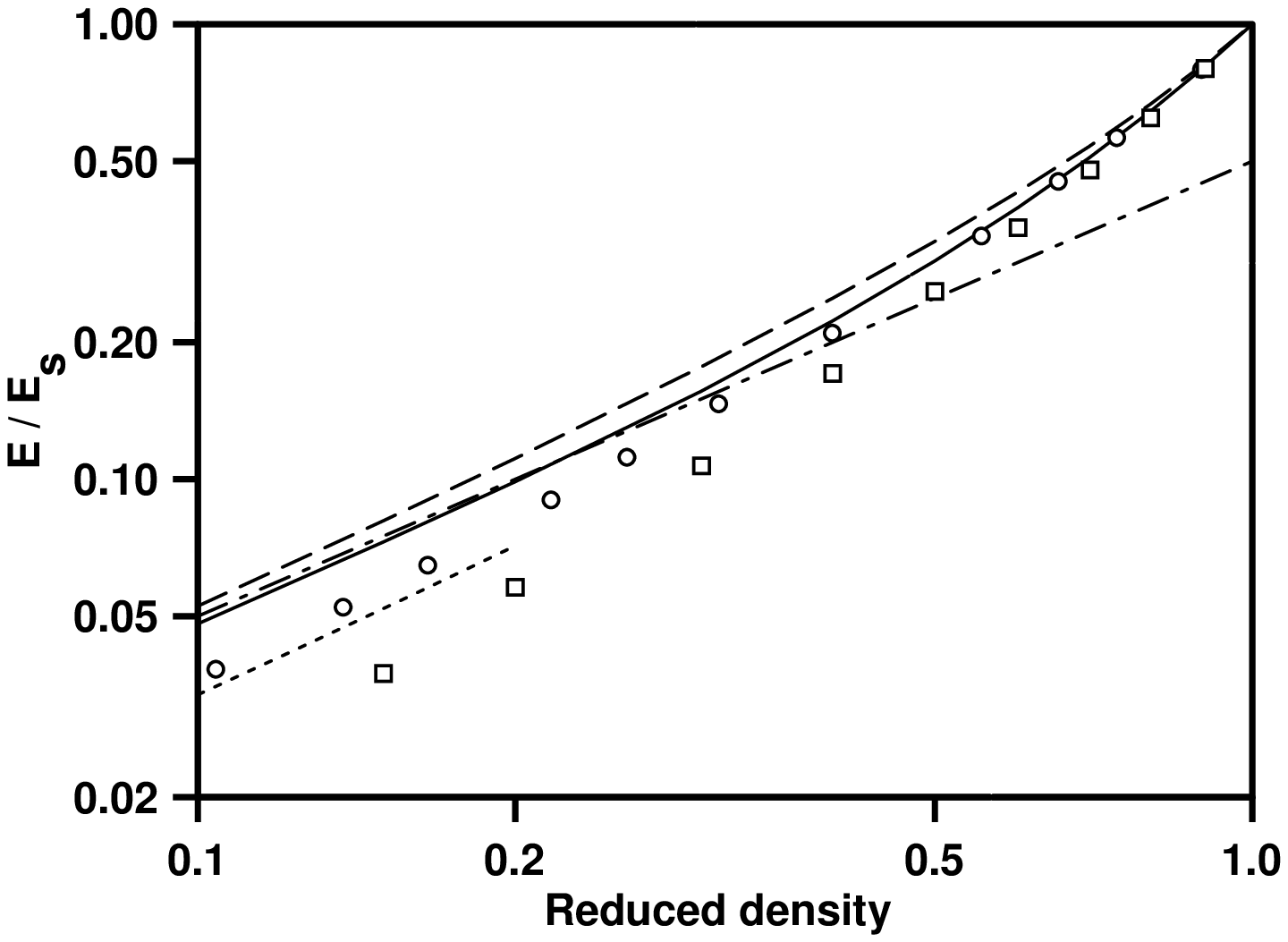}

Figure 10
\end{minipage}
\begin{minipage}[bt!]{.5\linewidth}
\centering \epsfxsize=.95\linewidth\epsfbox{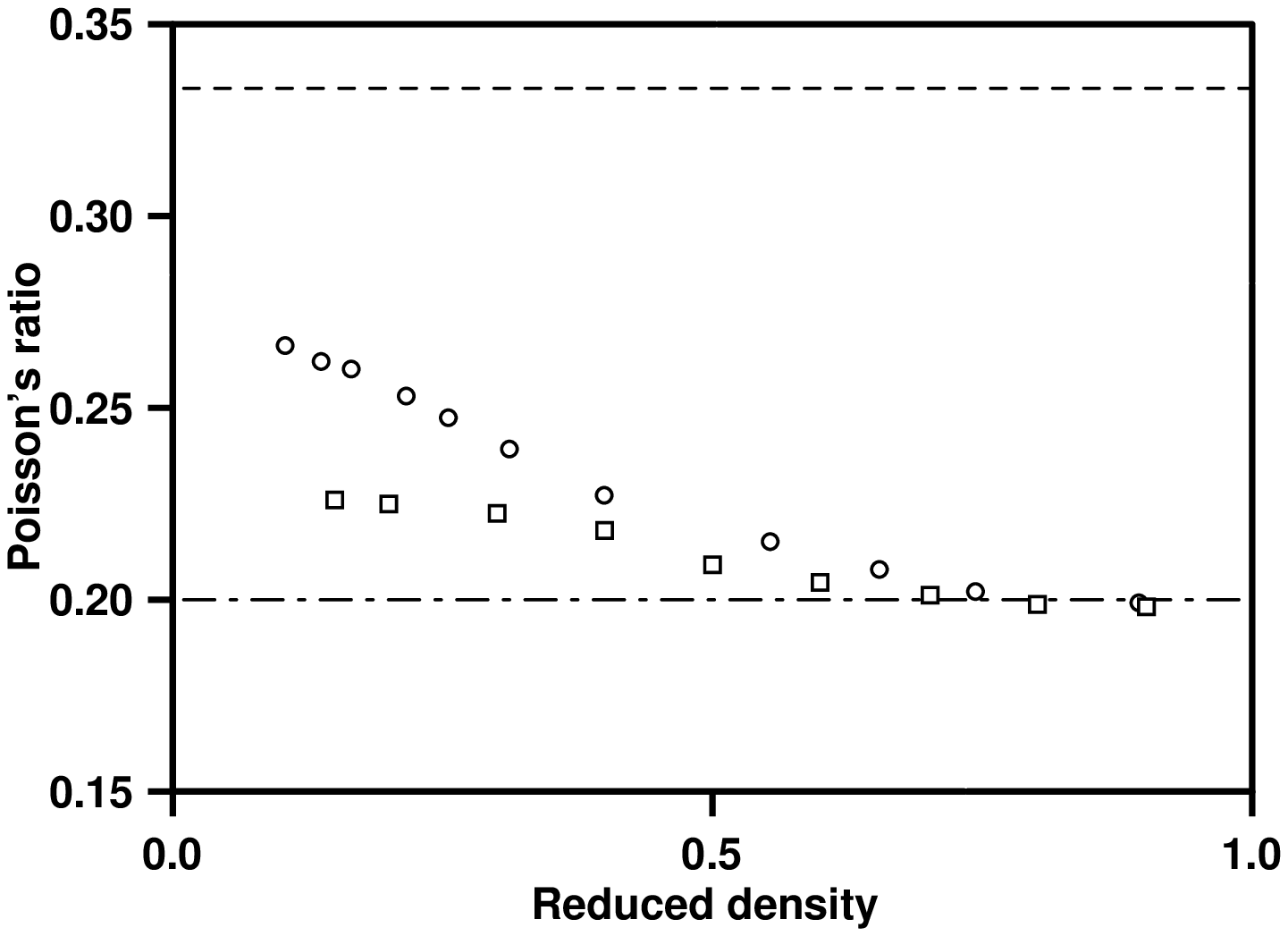}

Figure 11
\end{minipage}
\caption{Comparison of the FEM data (symbols) with theory (lines).
The data shown is for the closed-cell Voronoi tessellation ($\circ$) and
the closed-cell Gaussian random field model ($\Box$).
The theories are due to
Christensen~\protect\cite{ChristensenLow86} (-- $\cdot$ --) and
Simone and Gibson~\protect\cite{Simone98} ($\cdots$).
We also show the Hashin-Shtrikman bound for all isotropic
materials (--- ---) and the 3-point bound on the modulus
of the Gaussian random field model (------).
\label{cft_Eclos}}
\caption{Comparison of the FEM data for closed-cell foams (symbols)
with theory (lines). The data shown is for the closed-cell Voronoi
tessellation ($\circ$) and
the closed-cell Gaussian random field model ($\Box$).
The lines correspond to
Christensen's theory~\protect\cite{ChristensenLow86} (-- $\cdot$ --) and
the empirical result $\nu_s$=0.33 of Gibson and 
Ashby~\protect\cite{Gibson88} (-- -- --). 
\label{cft_Sclos}}
\end{minipage}
\end{figure}

In Fig.~\ref{cft_Eclos}, we compare FEM data
for the closed-cell foams with the prior results discussed in
section~\ref{prior}. The results for the random Voronoi tessellation are
seen to be around 10 \% greater than Simone and Gibson's results for
the tetrakaidecahedral model. It is not clear if this implies that the
disordered tessellation is stiffer than a regular tessellation.
Indeed, Grenestedt~\cite{Gren_shape99} found that a
tessellation of a randomly perturbed BCC lattice
was in fact 10 \% weaker than the tetrakaidecahedral model.
Therefore, it is possible that differences between the models are explicable
in terms of systematic discretization errors (which we estimate to be around
10 \%). The modulus of the Gaussian random field model
is considerably below the estimate for
the tetrakaidecahedral model. This can be attributed to the highly
irregular cell shapes and curved faces of the Gaussian model.
At low densities, Christensen's result significantly overestimates data
for both random models, indicating that the assumption of ``straight-through''
faces is not justified for random foams.

As expected, all of the data fall below the Hashin-Shtrikman bound for
isotropic materials. For the Gaussian random field model it is possible to
evaluate the 3-point statistical correlation function~\cite{Roberts96a},
and calculate the more restrictive 3-point upper bound~\cite{Milton82b}.
Figure~\ref{cft_Eclos} shows that the 3-point bound still does not provide
a good estimate of the Young's modulus and computation is necessary.

To apply the semi-empirical formula of Gibson and Ashby given in
Eq.~(\ref{thyGA_C}), we need to estimate the
fraction of mass contained in the cell edges $\phi$.
To determine $\phi$ for the Voronoi tessellation
we have deleted all the faces from the model and recorded the remaining
density (see Table~\ref{tabvt}). For $\rho/\rho_s < 0.3$, we find
$\phi<\frac12$ indicating that the prediction of Eq.~(\ref{thyGA_C}) is
in fact greater than the Hashin-Shtrikman bound.
For the closed cell Gaussian random field model,
an analytic result for $\phi$ can be derived as
follows. 
If, instead of a union set, we
form the intersection set of two level-cut random-fields,
we obtain an open-cell model comprised of the cell-edges of the
closed-cell model. Denoting the reduced density of the
open- and closed-cells Gaussian models as $p_{\rm op}$ and $p_{\rm cl}$,
it can be shown that
$p_{\rm cl}=\sqrt p_{\rm op}(2-\sqrt p_{\rm op})$~\cite{Roberts_TAg99}.
The fraction of mass in the edges is then
\begin{equation}
\phi=\frac{p_{\rm op}}{p_{\rm cl}}=\frac{(1-\sqrt{1-p_{\rm cl}})^2}{p_{\rm cl}}.
\end{equation}
Now for $p_{\rm cl} < \frac89$, $\phi < \frac12$, indicating that the
semi-empirical formula will exceed the Hashin-Shtrikman upper bound
at lower densities. Therefore the semi-empirical formula does not provide
an accurate estimate of the elastic modulus for either model.

The Poisson's ratio of the closed cell foams are compared with
predictions in Fig.~\ref{cft_Sclos}.
The FEM data for the closed cell Voronoi tessellation and Gaussian
random field models increase from 0.2 (the solid value) to about
0.24 and 0.28, respectively, as density decreases.
The results lie between the predictions
$\nu=0.2$ and $\nu=0.33$.  In a related study~\cite{GR_UP}, we have shown that
the Poisson's ratio becomes independent of the solid-Poisson's ratio
at low densities indicating that the predictions are not correct
for the models studied here.

\section{Comparison of FEM results with experiment}

To illustrate the utility of the FEM, we compare the computed results to
experimental data (it should be remembered that the computed results
have about a 10 \% uncertainty, mainly due to digital resolution). 
Since real foams can have densities lower
than those we are currently
able to computationally study, we use the formula
$E/E_s=C(\rho/\rho_s)^n$ to extrapolate the results.
This is justified by the fact that the low density FEM data appear
to fall on a straight line when plotted against log-log axes.
Accurate comparison of theoretical and experimental results is hindered
by the imprecision involved in estimating the properties of the solid skeleton
$E_s$ and $\rho_s$.  We report $E_s$ and $\rho_s$ when they have been given, but
some data sets are reported only in terms of $E/E_s$ and $\rho/\rho_s$.
Some of the data sets we have taken from the literature
have been previously summarized~\cite{Gibson88,Green85}.

\begin{figure}

\begin{minipage}[t!]{1.\linewidth}
\begin{minipage}[bt!]{.5\linewidth}
\centering \epsfxsize=.95\linewidth\epsfbox{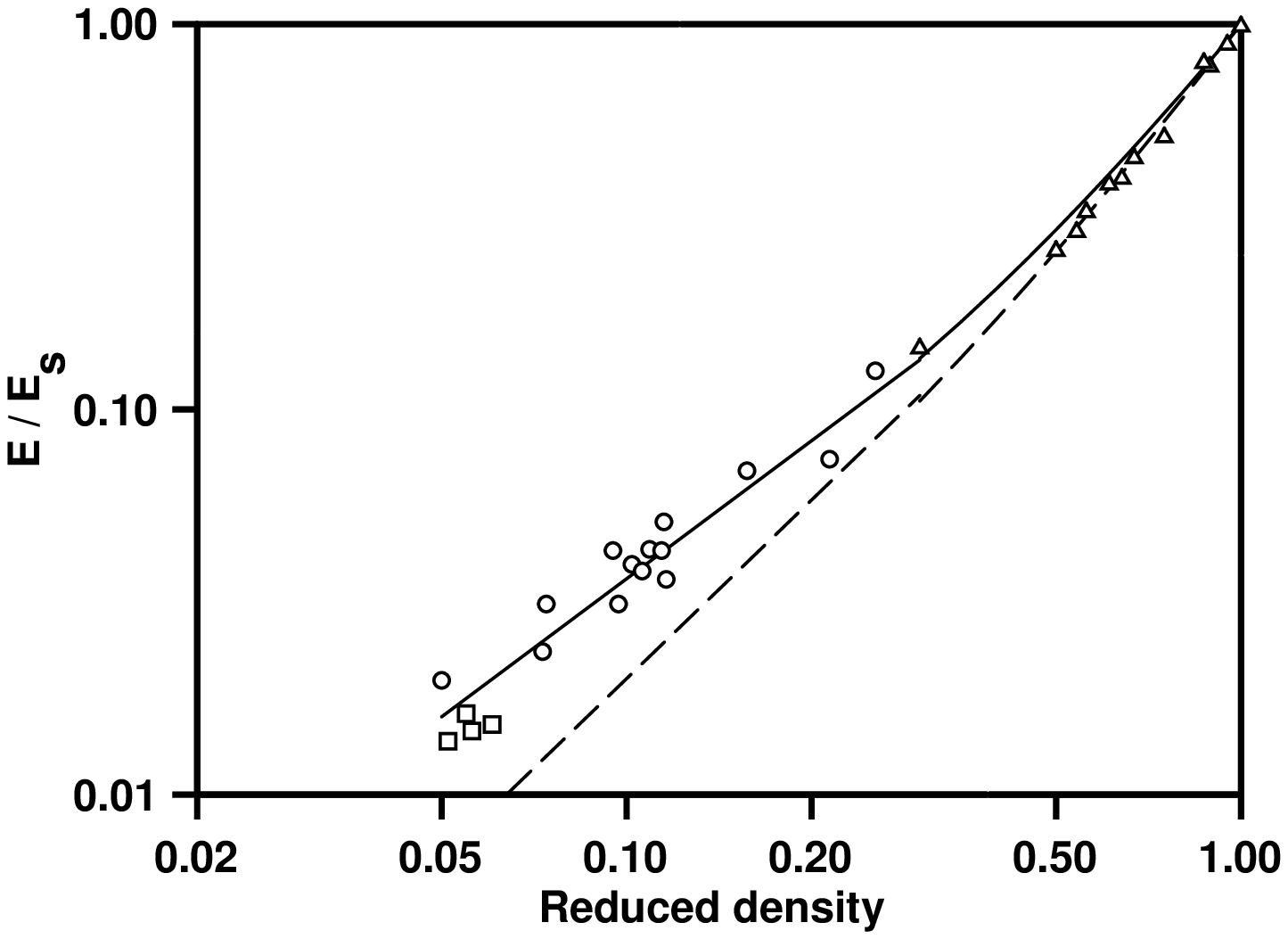}

Figure 12
\end{minipage}
\begin{minipage}[bt!]{.5\linewidth}
\centering \epsfxsize=.95\linewidth\epsfbox{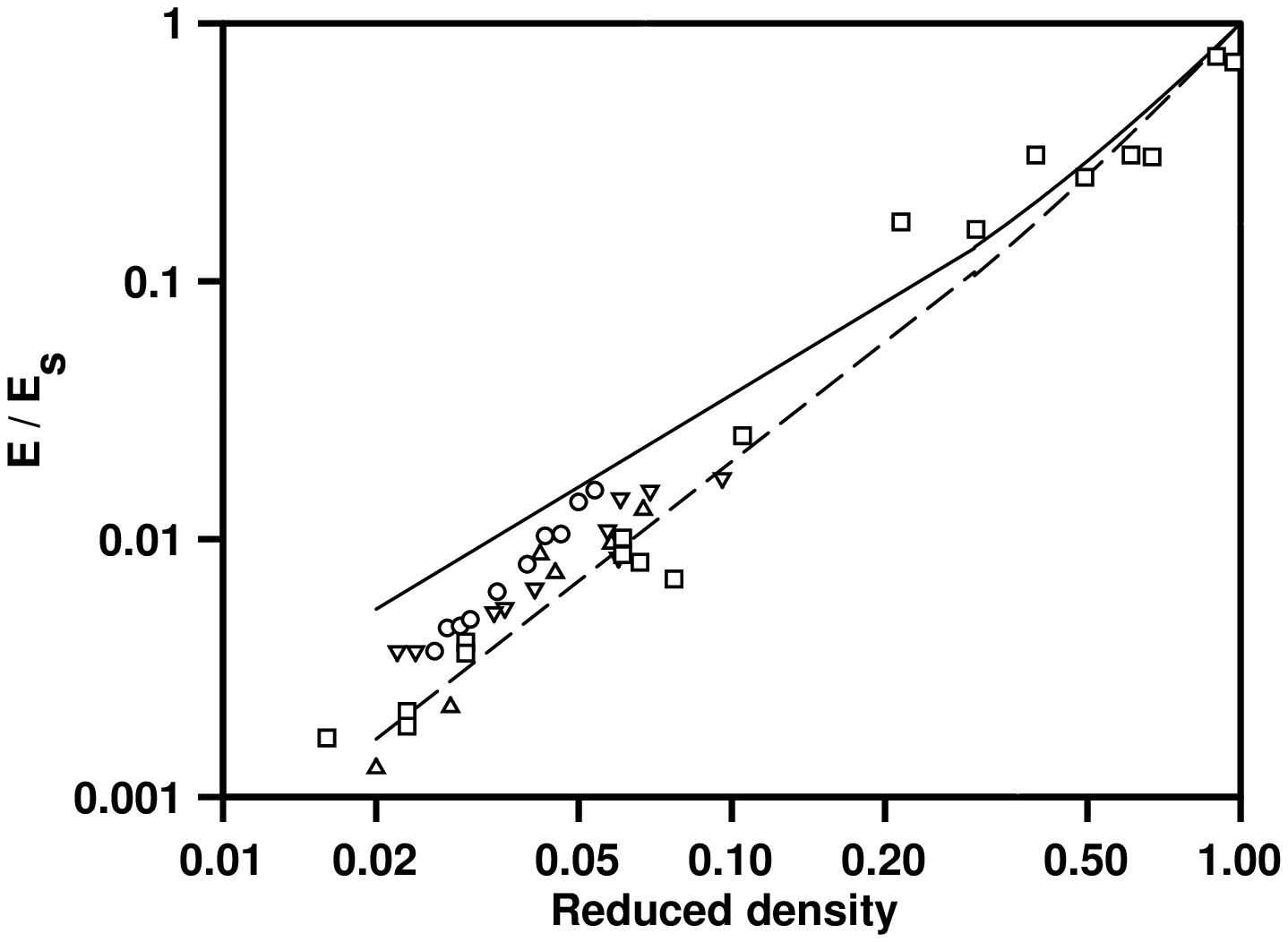}

Figure 13
\end{minipage}

\caption{Young's modulus of foamed glasses with closed cells. The
data is from Morgan {\em et al}~\protect\cite{Morgan_Glass81} ($\Box$),
Zwissler and Adams~\protect\cite{Zwissler83} ($\circ$)
($E_s$=69 GPa~\protect\cite{Green85}) and
Walsh {\em et al}~\protect\cite{Walsh_Glass65} ($\triangle$) ($E_s$=75 GPa).
The solid line
(------) corresponds to the closed-cell Voronoi tessellation.
Results for the closed-cell GRF model (-- -- --)
are shown for comparison. 
\label{cc_foamg}}

\caption{Young's modulus of closed cell polymer foams. The
data is for extruded polystyrene~\protect\cite{ChanNak69} 
($\circ$, $E_s$=1.4 GPa and $\rho_s$=1050 Kg/m$^3$),
polystyrene beads~\protect\cite{Mills_foam99}
($\triangle$, $E_s$=3.0 GPa and $\rho_s$=1100 Kg/m$^3$)
expanded polystyrene~\protect\cite{BaxterJones72}
($\Box$, $E_s$=2.65 GPa and $\rho_s$=1020 Kg/m$^3$)
and for low-density polyethylene~\protect\cite{Clutton91}
($\bigtriangledown$).
The solid (------)
and dashed lines (-- -- --) correspond to the closed cell Voronoi
tessellation and Gaussian random field models. 
\label{cc_poly}}

\end{minipage}
\end{figure}

Data for closed cell porous
glass~\cite{Morgan_Glass81,Zwissler83} (Fig.~\ref{cc_foamg})
agrees well with the FEM results obtained using the closed cell
Voronoi tessellation. Micrographs of the glass studied by
Zwissler and Adams~\cite{Zwissler83} indicate a structure
similar to that of the Voronoi tessellation shown in Fig.~\ref{vt3D},
indicating that the model is appropriate.
Data for closed cell polymer foams is shown in 
Fig.~\ref{cc_poly}. The data for expanded
polystyrene~\cite{BaxterJones72} generally
agree with the predictions of the closed-cell Gaussian random field model.
The data for extruded polystyrene~\cite{ChanNak69}
decreases from the Voronoi tessellation towards the Gaussian random field
result as the density decreases. Micrographs of
polystyrene~\cite{Roberts95b} indicate a cell structure similar to
that of the Voronoi tessellation, but the cell walls show some curvature.
This may explain why the results for the random field model
(which contains curved cell walls) more closely matches the data.

\section{Discussion and Conclusion}

We have used the finite element method to estimate the Young's modulus
of realistic random models of isotropic cellular solids. 
At low densities, the results can be described by the scaling
relation $E/E_s=C(\rho/\rho_s)^n$, where the parameters are reported in
the text.
At moderate to high densities, the results were described by
Eq.~(\ref{percfit}).
The equations used to describe the data are chosen to
provide a reasonable fit, and consequently the fitting parameters do
not have clear physical significance. It would be ideal
to resolve the modulus into components from edge-bending and
plate-stretching. However, the non-linear interaction between arbitrarily
shaped cells and deformation makes this task impossible
in all but the simplest of models.

All our results were obtained using a solid
Poisson's ratio of $\nu_s=0.2$. It has recently been shown~\cite{GR_UP}
that the Young's modulus only varies by around 2 \% for $0 < \nu_s < 0.5$
indicating that our results are valid for all usual values of the
solid Poisson's ratio.
The fitting relations we have derived
can be used to predict the properties of cellular materials
that have a microstructure similar to one of the models,
and can be useful for interpreting experimental data.

Our results for closed-cell Voronoi tessellations were in general
agreement with earlier studies on the
tetrakaidecahedral
foam, i.e.,
$E/E_s \approx \frac13 (\rho/\rho_s)$ as $(\rho/\rho_s) \to 0$~\cite{Renz82}. 
The actual exponent was $n=1.19$ which is greater than the 
value $n=1$ obtained for single-cell models and scaling
arguments [Eq.~(\ref{thyGA_C}), $\phi < 1$].
If more than 70 \% of of the
cell faces are removed, the Young's modulus exponent 
increased to $n=2$,  indicating that edge bending becomes the
dominant mechanism of deformation. 

The closed-cell random field model, with curved cell walls, showed
a significantly greater density dependence than the Voronoi tessellation
(with exponent $n\approx 1.5$). Although
the minimum density at which we were able to measure properties was
$(\rho/\rho_s)\approx0.15$, the data showed no evidence of
adopting a linear decay for $(\rho/\rho_s)\to 0$,
as suggested by theory.

The semi-empirical closed-cell theory given in
Eq.~(\ref{thyGA_C}) was found to not be applicable to 
the models studied here, since the low
edge fractions ($\phi < 0.5$) caused the equation to exceed known
upper bounds on the Young's modulus.
Moreover, attempts to describe the deformation of closed or partially
closed cellular materials in terms of a bending (exponent $n=2$)
and plate stretching (exponent $n=1$) component were not successful.
Instead, our results indicate that these mechanisms combine non-linearly
and are best represented by a non-integer power law.
Variational upper bounds, and other predictions for random closed
cell foams, were found to significantly over-estimate the 
Young's moduli of the models. Therefore numerical simulation
must be relied on for accurate predictions.

In this study, we have shown that it is important to
consider large-scale (multi-cellular) models of random cellular solids in
order to obtain realistic elastic properties.  While the modulus of
the closed cell Voronoi tessellation can be approximately described
by a single cell of the 
tetrakaidecahedral model, it is not possible to model the effect
of missing faces and irregular cells with curved walls using
single-cell models.
Our results are consistent with experimental data, and show a more
complex density dependence than predicted by conventional theories
based on scaling arguments and periodic cell models.
Our results focus on the effect of multi-cellular disorder, rather than local
characteristics (e.g., distribution of mass between cell edges and walls) of
cellular materials, for the following reasons. First, it is difficult to 
simultaneously
model the local and global variables with finite computational power,
and second, study of single cell models probably provides a more
fruitful route to understanding the influence of
local cell-character on the overall properties.
We believe that the results of both approaches may be
beneficially combined.

\vspace{4mm}

{\small \noindent {\em Acknowledgements}---
A.R.\ thanks the Fulbright Foundation and Australian Research Council
for financial support. We also thank the Partnership for
High-Performance Concrete program of the National Institute
of Standards and Technology for partial support of this work.
}


\end{document}